**Соколов Володимир Юрійович**
старший викладач кафедри інформаційної та кібернетичної безпеки
Київський університет імені Бориса Грінченка, Київ, Україна
OrcID: 0000-0002-9349-7946
vladimir.y.sokolov@gmail.com

# ПОРІВНЯННЯ МОЖЛИВИХ ПІДХОДІВ ЩОДО РОЗРОБКИ НИЗЬКОБЮДЖЕТНИХ АНАЛІЗАТОРІВ СПЕКТРУ ДЛЯ СЕНСОРНИХ МЕРЕЖ ДІАПАЗОНУ 2,4–2,5 ГГЦ

**Анотація.** В статті приведені розробка, реалізація і дослідження роботи аналізаторів спектру, які можуть бути використанні у сенсорних мережах і системах інтернету речей. В якості робочого частотного діапазону вибраний ISM — 2,4–2,5 ГГц. На етапі вибору апаратного забезпечення проведений порівняльний аналіз існуючих доступних мікроконтролерів для аналізу спектру, вибір апаратних інтерфейсів, замовлення необхідних модулів і електричних компонентів, а також вхідних контроль. Підчас розробки реалізовано кілька варіантів аналізаторів спектру на базі мікроконтролера і мікрозбірки TI Chipcon CC2500 з USB-інтерфейсами, а також модулів Cypress CYWUSB6935 з LPT- і USB-інтерфейсами. На етапі розробки проведено розробку друкованої плати, її виготовлення, монтаж компонентів, програмування мікроконтролерів, перевірка роботоздатності збірки, внесення виправлень, під'єднання до персонального комп'ютера і збірка в корпусі. Також проведений аналіз існуючого програмного забезпечення для збору інформації про стан безпроводового ефіру. За результатами порівняльних дослідів різних збірок аналізаторів спектру отримані спектрограми для різних типів сигналів. За цими типовими спектрограмами проведений порівняльний аналіз роботи різних прототипів. Запропоновані підходи до побудови сенсорів на базі аналізаторів спектру дозволяють створювати малопотужні модулі для вбудовування в існуючі безпроводові інформаційні мережі підприємств для запобігання міжканальної інтерференції і забезпечення цілісності передавання даних. В результаті експериментів видно, що не всі існуючі модулі сильно відрізняються за характеристиками, а якість їх роботи напряму пов'язана з типом і якістю антени. В статті приведено принципові електричні схеми, перелік елементів, приклади друкованих плат, програматори, програмне забезпечення і прототипи.

**Ключові слова:** аналіз спектру; безпроводова мережа; трансивер; ISM-діапазон.

## 1. ВСТУП

В Україні все більше нових домашніх і промислових мереж будується з повним чи частковим використанням безпроводових технологій. За останні кілька років такі технології стали стандартом де-факто. Кількість мереж збільшується через їх цінову доступність і спрощення роботи з ними, появі промислових систем з роумінгом, широкий вибір антенного обладнання і дозволи використання, закріплені на законодавчому рівні.

У статті розглянуте головне питання забезпечення безпеки передавання інформації у безпроводових системах: її доступність. При проектуванні безпроводової мережі не можливо передбачити всіх нюансів: перевідбиття, затінення, спрямованість антен приймачів тощо, тому після побудови реальної системи потрібно її перевірити і зменшити вплив негативних факторів. Аналізатори спектру допомагають вирішити цю проблему.

Аналізатори спектру не збільшують доступність інформації в мережі, але сприяють її вдосконаленню, через виявлення слабких місць, заповненості спектру іншими мережами (запобігання колізій/інтерференції). На відміну від стандартних засобів, доступних в мережевих картах, аналізатор спектру збирає інформацію про рівень шуму, наприклад, від магнетронів (у мікрохвильових печах), може виявити стаціонарні завади і роботу іншого обладнання (Bluetooth, ZigBee, радіотелефонів, дитячих іграшок, трансиверів відеосигналів, медичних датчиків тощо). Також аналізатор спектру «бачить» не тільки службові пакети, за якими оцінюється рівень сигналу в мережевих картах, а і усі інші пакети. Тому тема статті є важливою і вкрай актуальною.

Попередні версії промислових аналізаторів спектру Pololu Wixel використовувалися авторами в якості сенсорів при проведенні досліджень антенної техніки [1] і побудові сенсорних мереж [2].

## 2. ПОРІВНЯННЯ РАДІОЧАСТОТНИХ ТРАНСИВЕРІВ

Аналізатор спектру — прилад для сканування та аналізу певного діапазону частот, що дозволяє виявляти в радіоефірі доступні безпроводові мережі зв'язку, точки доступу та клієнтські пристрої, а також отримувати дані про їхні характеристики. Так, наприклад, використання сканера частот дозволяє ідентифікувати назви виявлених радіомереж, їх рівень сигналу, використовуваний тип шифрування даних та інші параметри. Найбільш часто спектральні Wi-Fi аналізатори можна зустріти в арсеналі фахівців, чия професійна діяльність пов'язана з розгортанням і настроюванням безпроводових мереж передачі даних стандартів 802.11 a/b/g/n/ac.

Використання сканера частот на етапі побудови Wi-Fi мережі дає можливість інженерові в короткий термін визначити наявність і рівень перешкод в радіоефірі, вибрати оптимальну частоту для роботи WLAN, а також розрахувати кількість і розташування точок доступу для забезпечення повної зони покриття. При виникненні збоїв в роботі мережі аналізатор спектру радіочастот здатний надати допомогу при проведенні діагностики та виявленні причини неполадки, в тому числі оцінити завантаженість мережі, виявити несанкціоновані підключення і пристрої, ідентифікувати і локалізувати радіоперешкоди в каналах [3], [4].

В даній роботі вибрані лише мікросхеми з низькою собівартістю, щоб їх можна було б використовувати в сенсорних мережах таких, як приведені в [2]. В табл. 1 зведені дані про шість найбільш вдалих трансиверів даного класу від виробників Nordic, Texas Instruments і Cypress.

*Таблиця 1*

**Основні характеристики трансиверів**

| Мікросхема | Частотний діапазон, МГц | Роздільна здатність, кГц | Діапазон потужності, дБмВт | Роздільна здатність, дБмВт |
|---|---|---|---|---|
| Nordic nRF24L01 [5] | 2400–2525 | 977 | –(85÷42) | 1,0 |
| TI Chipcon CC2500 [6] | 2400–2483,5 | 58–812 | –(104÷13) | 0,8 |
| TI Chipcon CC2511-F32 [7] | 2400–2483,5 | 58–812 | –(110÷6,5) | 0,5 |
| Cypress CYRF6934 [8] | 2400–2483 | 1000 | –(90÷40) | ~4,1 |
| Cypress CYRF6935 [9] | 2400–2483 | 1000 | –(95÷40) | ~3,1 |
| Cypress CYRF6936 [10] | 2400–2497 | 1000 | –(97÷47) | ~1,3 |

З порівняння видно, що найбільш вдалими є трансивери фірми Texas Instruments. Далі приведені досліди для порівняння роботи трасиверів тільки двох фірм (Texas Instruments і Cypress), так як модуль Nordic nRF24L01 має надто малий діапазон виміру потужності сигналу і через це межі його застосування досить малі.

В статті розглянуті чотири апаратні реалізації аналізаторів спектру на базі:
– мікроконтролера CC2500 (з USB-інтерфейсом);
– модуля CC2500 (USB);
– модуля CYWUSB6935 (LPT);
– модуля CYWUSB6935 (USB).

**3. АПАРАТНИЙ АНАЛІЗАТОР НА БАЗІ МІКРОКОНТРОЛЕРУ CC2500**

Для аналізу цілісності передавання даних ми будемо використовувати апаратний аналізатор спектру, який побудований на радіочастотному трансивері, а саме Chipcon CC2500. Схема пристрою показана на рис. 1, а перелік електронних компонентів — в табл. 2.

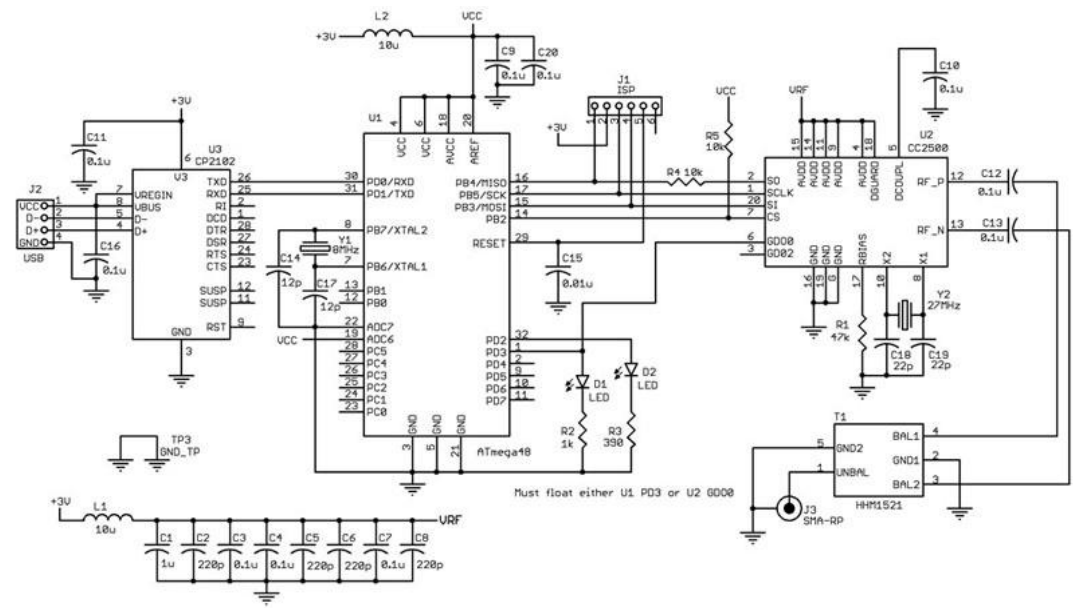

*Рис. 1. Електрична принципова схема аналізатора спектру на Chipcon CC2500*

Для початку було розроблено друковану плату згідно схеми, приведеної на рис. 1. Приклад такої двосторонньої плати приведено на рис. 2а. Цей проект переноситься на термопапері, який прасується до того моменту, поки колір з білого не зміниться на сірий. Після цього беремо цей термопапір і за допомогою лазерного принтера друкуємо на ньому плату. На рис. 2б продемонстровано термопапір з надрукованою платою. Плата виконана на склотекстоліті розмірами 96×71 мм і товщиною 1 мм. Роздруковані плати прикладаються до склотекстоліту та переносяться праскою близько 5 хвилин з кожного боку. Плата охолоджується і переноситься в ємність з теплою водою, а за півгодини папір видаляється.

*Таблиця 2*

**Список електронних компонентів до аналізатора спектру на Chipcon CC2500**

| Позначення | Значення | Опис | Кількість |
|---|---|---|---|
| C1 | 1 мкФ | конденсатор | 1 |
| C2, C5, C6, C8 | 220 пФ | конденсатор | 4 |
| C3, C4, C7, C9–C11, C16, C20 | 100 нФ | конденсатор | 8 |
| C12, C13 | 100 пФ | конденсатор | 2 |
| C14, C17 | 12 пФ | конденсатор | 2 |
| C15 | 10 нФ | конденсатор | 1 |
| C18, C19 | 22 пФ | конденсатор | 2 |
| J1 | ISP | 6-контактний з'єднувач | 1 |
| J2 | USB | USB-роз'єм | 1 |
| J3 | SMA | SMA-з'єднувач | 1 |
| L2, L1 | 10 мкГн | індуктивність | 2 |
| R1 | 47 кОм | резистор | 1 |
| R2 | 1 кОм | резистор | 1 |
| R3 | 390 Ом | резистор | 1 |
| R4, R5 | 10 кОм | резистор | 2 |
| T1 | HHM1521 | трансформатор 2,4 ГГц | 1 |
| U1 | ATmega48 | ATmega48 TQFP | 1 |
| U2 | CC2500 | CC2500 трансивер | 1 |
| U3 | CP2102 | USB-UART контролер | 1 |
| Y1 | 8 МГц | кварц | 1 |
| Y2 | 27 МГц | кварц | 1 |

Плату переноситься в ємність з хлорним залізом, і плата витравлюється. (З першої спроби не вийшло зробити через тонкі доріжки коло трансивера, які почали «відлітати» відразу ж при травленні плати). За допомогою наждачного паперу обробляється заготовка, просверлюються отвори за допомогою свердл діаметром 0,5 і 2 мм. В результаті отримуємо готову плату для монтування елементів (рис. 2в).

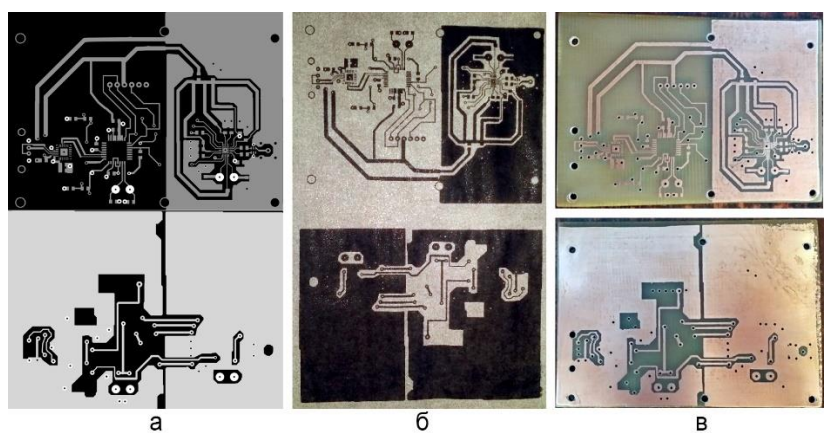

*Рис. 2. Друкована плата аналізатора спектру на Chipcon CC2500:*
*а — проект в графічному редакторі; б — термопапір з надрукованою платою;*
*в — витравлена плата*

Після монтажу елементів проводиться програмування мікроконтролера Atmega48 за допомогою USB-програматора для AVR (на рис. 3) [12], [13].

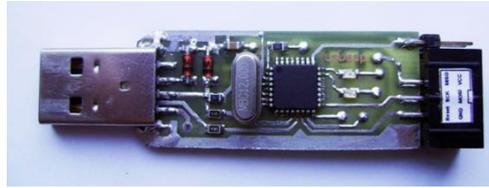

*Рис. 3. Програматор для AVR мікроконтролерів*

Для програмування нам необхідно з'єднати шість проводів від програматора з контактними майданчиками на платі. Програматор не побачив мікроконтролер, тому було вирішено випаяти мікроконтролер і підпаяти безпосередньо до ніжок мікросхеми. На рисунку 4а можна побачити вже готову плату, але з випаяним мікроконтролером.

Після прошивки мікроконтролера припаюємо його назад на плату і підключаємо до комп'ютера, встановлюємо драйвери для пристрою і запускаємо програму для аналізу спектру, але при першому запуску пристрій не запрацював. Перевірили всі доріжки на платі за допомогою мультиметра і було виявлено, що на одній доріжці був обрив, він був успішно усунутий за допомогою паяльника й олова. Після чого плата була встановлена в корпус (див. рис. 4б)

Проводимо ряд тестів і спостерігаємо. Якщо піднести руку або інший предмет до пристрою, створюється наводка, тому було нами вирішено поставити екран для зменшення завад в області, де розміщається трансивер. Екран ми виготовили з міді (рис. 4в).

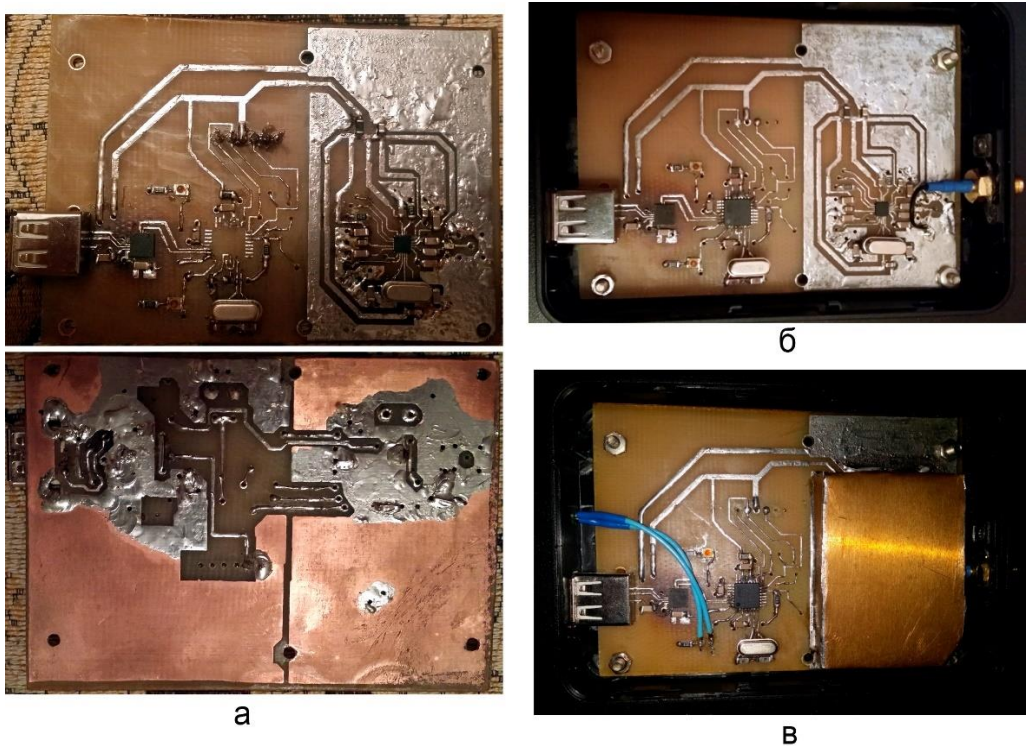

*Рис. 4. Друкована плата аналізатора спектру на Chipcon CC2500:*
*а — із змонтованими елементами; б — встановлена в корпус; в — з припаяним екраном*

Результати роботи аналізатора спектру представлені на рис. 5.

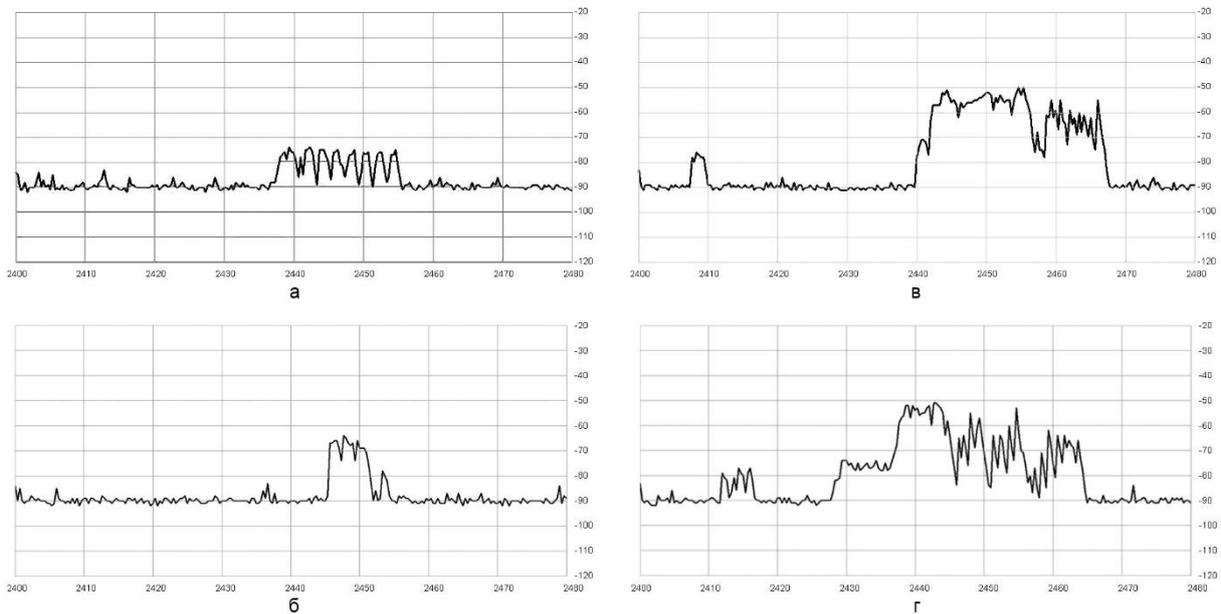

*Рис. 5. Аналіз спектру, отриманого від Chipcon CC2500:*
*а — без використання антени і екрану; б — з використанням екрану, але без антени;*
*в — з використанням антени, але без екрану; г — з використанням антени і екрану*

Підчас проведення експериментів згорів мікроконтролер через недостатній тепловідвід, тому було прийнято рішення витравити нову плату, замінити мікроконтролер і антену.

## 4. АПАРАТНИЙ АНАЛІЗАТОР НА БАЗІ МОДУЛЮ CC2500

Для реалізації аналізатора спектру зі змінним трансивером, нам необхідний сам трансивер, який ми замовили з Китаю. Він має компактні розміри. На рисунку 6 показано, який він має вигляд.

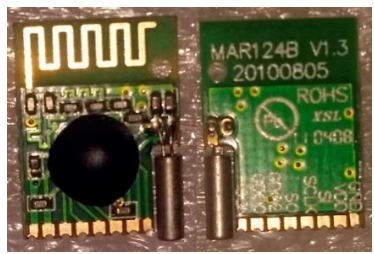

*Рис. 6. Зовнішні вигляд модуля трансивера CC2500*

Друкована плата була змінена. На рису. 7а показано, як тепер виглядає друкована плата. Цього разу не використовувався термопапір, а фотопапір. Далі беремо фотопапір і за допомогою лазерного принтера друкуємо на ній друковану плату.

Беремо склотекстоліт з розмірами 60×50 мм товщиною 1 мм і прикладаємо наші роздруковані плати, прикладаємо до склотекстоліту та гладимо праскою приблизно 5 хвилин з кожного боку. Даємо охолонути і кладемо в миску з теплою водою. Потім десь через 30 хвилин пальцем видаляємо папір. На рис. 7б показані плати після видалення паперу. Запаюємо всі деталі, програмуємо мікроконтролер Atmega48 та підпаюємо модуль, все заливаємо рідкою гумою (див. рис. 7в).

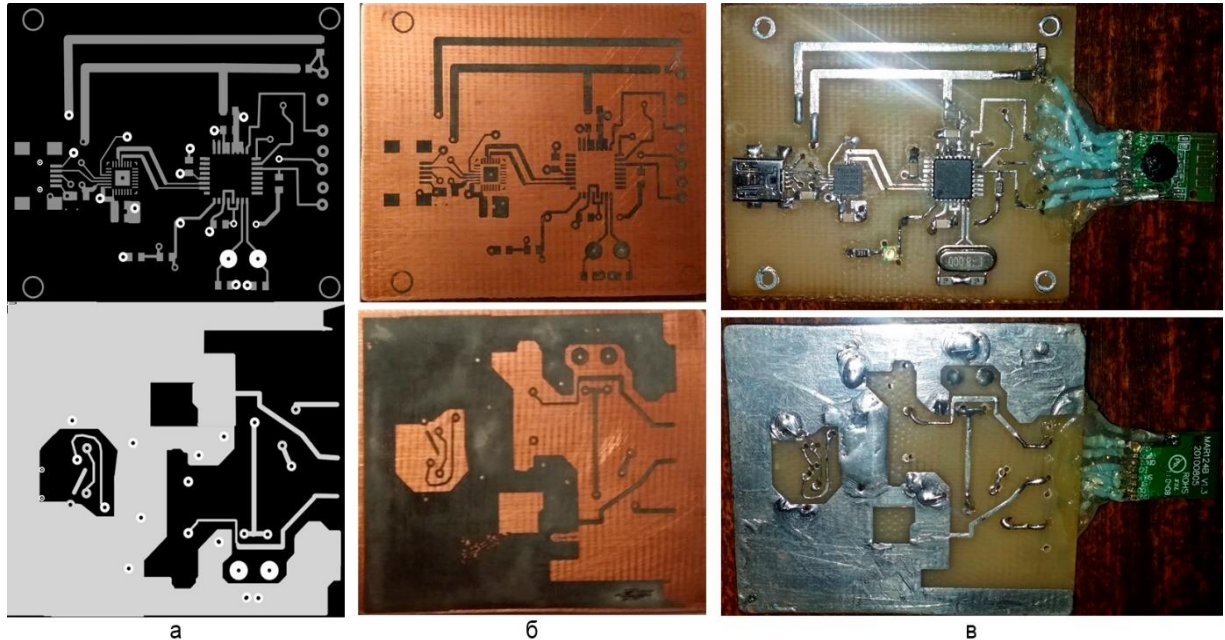

*Рис. 7. Друкована плата аналізатора спектру на модулі CC2500:*
*а — проект в графічному редакторі; б — витравлена; в — у готовому пристрою*

## 5. АПАРАТНИЙ LPT-АНАЛІЗАТОР НА БАЗІ МОДУЛЮ CYWUSB6935

Для реалізації аналізатора спектру за супергетеродинним принципом використана мікрозбірка Cypress CYWUSB6935 [9]. На рис. 8 показана принципова схема, в табл. 3 — перелік компонентів, а на рис. 9 — зібраний на монтажній платі перехідник.

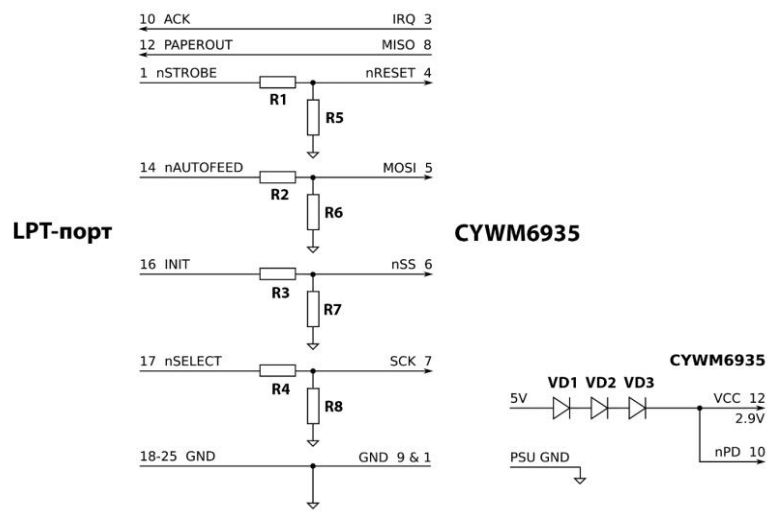

*Рис. 8. Електрична принципова схема під'єднання до модуля CYWUSB6935*

*Таблиця 3*

**Список електронних компонентів до USB-аналізатора на модулі CYWUSB6935**

| Позначення | Значення | Опис | Кількість |
|---|---|---|---|
| R1–R4 | 10 кОм | резистор | 1 |
| R5–R8 | 15 кОм | резистор | 4 |
| VD1–VD3 | 1N4001 | діод | 3 |

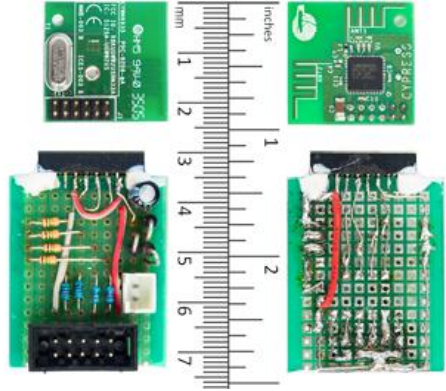

*Рис. 9. Зовнішній вигляд готового LPT-аналізатора спектру на модулі CYWUSB6935*

При зчитуванні напряму з LPT-порта отримані данні в наступному вигляді (частина пакетів з рівнем RSSI):

```
frame: [0,0,1,1,0,0,2,1,0,2,0,0,4,31,30,29,0,0,0,0,0,3,0,0,0,0,0,0,0,
0,0,0,1,1,0,0,0,1,0,2,0,0,0,0,0,0,0,1,0,0,2,0,0,0,2,0,0,0,1,0,4,5,1,1,
0,0,1,1,0,1,1,1,0,0,0,0,3,2,1,0,0,0,0,1,]
```

Після збору даних отримано результуючу картинку спектру, показану на рис. 10.

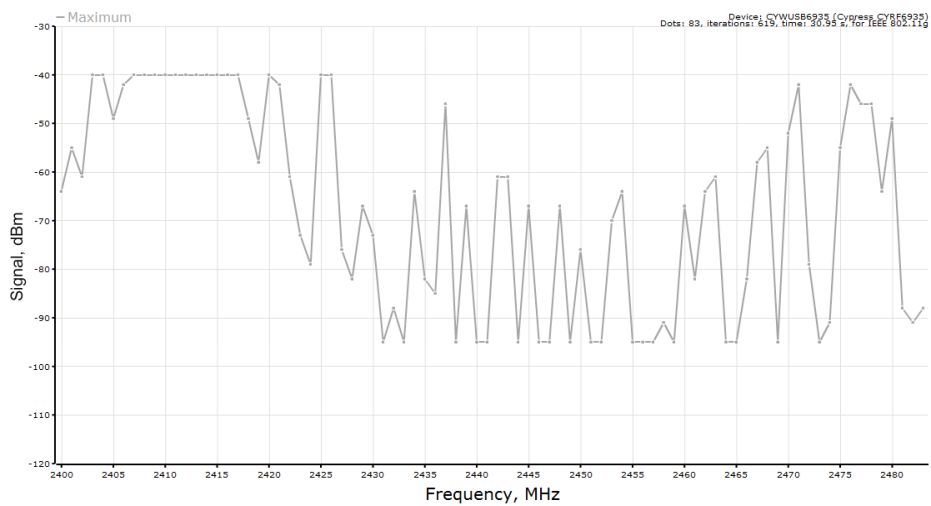

*Рис. 10. Результат аналізу даних від аналізатора спектру на модулі CYWUSB6935*

Конструкція має суттєвий недолік — робота на застарілому LPT-інтерфейсі, який нечасто зустрічається в сучасних комп'ютерах. Крім того, потрібне додаткове живлення. В даному випадку застосовувалось живлення від USB-порта послідовно підключеними трьома діодами для зниження напруги.

## 6. АПАРАТНИЙ USB-АНАЛІЗАТОР НА БАЗІ МОДУЛЮ CYWUSB6935

Принципова схема аналізатора спектру зображена на рис. 11. Оскільки мікроконтролер має в собі все необхідне для USB, в тому числі стабілізатор напруги 3,3 В, буфер пам'яті і приймач, все що потрібно зробити — підключити USB-кабель до виходів 15 і 16 мікроконтролера і конденсатор до виводу 14 для фільтрації напруги 3,3 В від вбудованого стабілізатора.

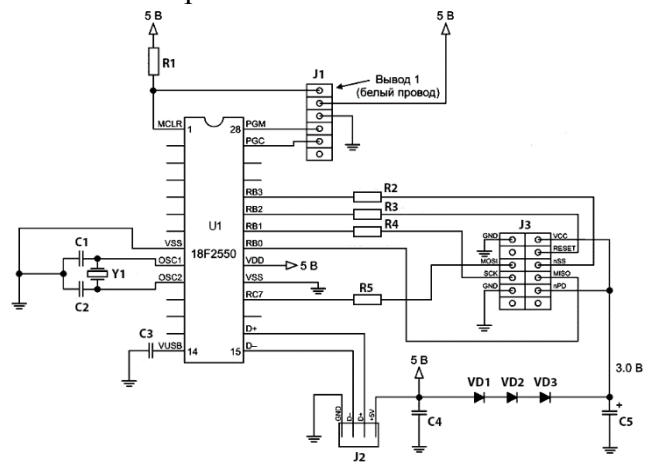

*Рис. 11. Електрична принципова схема керування модулем CYWUSB6935*

Тактування мікроконтролера здійснюється від кварцового резонатора 20 МГц з двома навантажувальними конденсаторами 15 пФ. Внутрішній дільник мікроконтролера ділить тактову частоту на 5, щоб отримати значення частоти 4 МГц, яка буде використовуватися для фазового автопідлаштування частоти, що працює на частоті 48 МГц. Це основна тактова частота, на якій працюють USB інтерфейс і ядро.

Резистор номіналом 10 кОм, підключений до висновку 1 мікроконтролера, підтягує висновок MCLR (скидання) до високого рівня.

Живлення сканер отримує від інтерфейсу USB, так як схема споживає незначний струм. Для живлення радіомодуля необхідно напруга від 2,7 до 3,6 В. Напруга порядку 3,0 В ми можемо отримати від шини 5 В, включивши послідовно 3 діода типу IN4001 (на кожному діоді падіння напруги близько 0,7 В). Це, звичайно ж, найпростіший і дешевий, але цілком надійний спосіб.

CYWUSB6935 має на входах захисні діоди. Це означає, що для керування можна використовувати 5-вольтові логічні сигнали мікроконтролера, включивши послідовні резистори для обмеження струму. Ми вибрали резистори з опором 3,3 кОм. Перелік компонентів представлений в табл. 4.

*Таблиця 4*

**Список електронних компонентів до LPT-аналізатора на модулі CYWUSB6935**

| Позначення | Значення | Опис | Кількість |
|---|---|---|---|
| C1, C2 | 15 пФ | конденсатор | 2 |
| C3 | 220 нФ | конденсатор | 1 |
| C4 | 100 нФ | конденсатор | 1 |
| C5 | 100 мкФ | конденсатор | 1 |
| J1 | ISP | 6-контактний з'єднувач | 1 |
| J2 | USB | USB-роз'єм | 1 |
| J3 | ISP | 10-контактний з'єднувач | 1 |
| R1 | 10 кОм | резистор | 1 |
| R2–R5 | 3,3 кОм | резистор | 4 |
| U1 | PIC18F2550-I/SP | мікроконтролер | 1 |
| VD1–VD3 | 1N4001 | діод | 3 |
| Y1 | 20 МГц | кварц | 1 |

Так як наша схема не занадто складна, для складання пристрою був обраний найпростіший шлях: макетна плата (див. рис. 12). Для підключення радіомодуля використовувався спеціальний багатопіновий роз'єм. Був використаний стандартний USB-роз'єм [13].

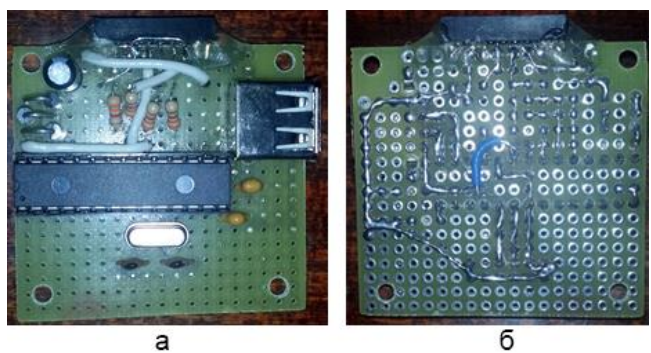

*Рис. 12. Готовий пристрій*

Програмуємо мікроконтролер PIC18F2550-I/SP через програматор (рис. 14).

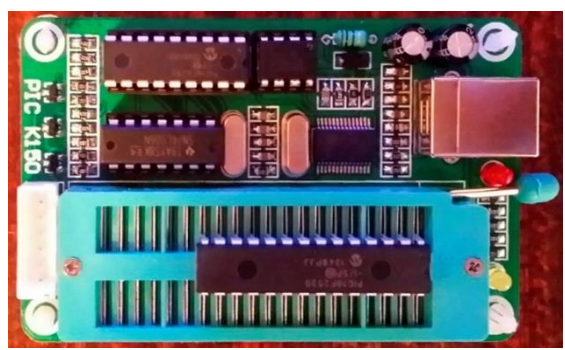

*Рис. 13. Програмування мікроконтролера*

## 7. АНАЛІЗ ПРОГРАМНОГО ЗАБЕЗПЕЧЕННЯ

Програма аналізатору спектру Low-Cost Spectrum Analyzer (LCSA) — це 32-бітний додаток для ОС Windows XP (не вимагає встановлення). При роботі з ОС Windows 7 потрібно встановити VP Windows XP Mode, щоб усунути проблему з драйверами. А перед використанням програмного забезпечення вперше, необхідно встановити драйвер для мікросхеми адаптера COM2USB Selicon Labs CP2102.

Після підключення аналізатора спектру, встановлення драйвера для CP2102 і запуску LCSA, програмне забезпечення повинно негайно розпочати збір даних і в режимі реального часу демонструвати ISM-діапазон спектру на 2,4 ГГц.

На вкладці *View* доступні такі функції: утримання піку, заможування піку, а також вибір фонів (чорного або білого). Також на вкладці *Plotting Options* є функції для налаштування амплітуди і частоти.

При роботі з віртуальними COM-портам може виникнути ситуація, коли потрібно задавати вручну номер порта в реєстрі системи:

```
REGEDIT4
[HKEY_LOCAL_MACHINE\HARDWARE\DEVICEMAP\SERIALCOMM]
"\\device\\slabser0"="COM20"
```

Весь огляд програмного забезпечення продемонстрований на рис. 14.

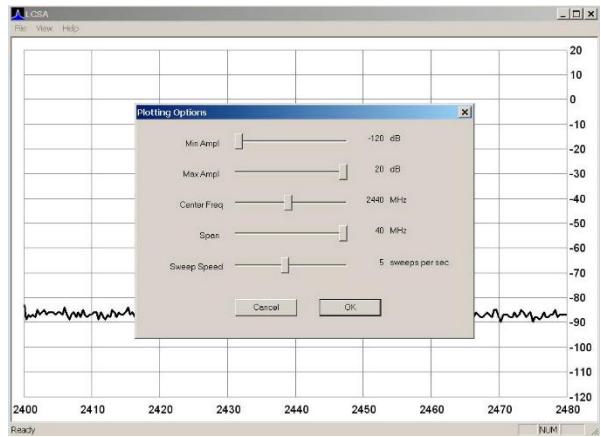

*Рис. 14. Головні функції програмного забезпечення LCSA*

Використовується стандартний USB-інтерфейс і може підключатися до будь-якого комп'ютера, на якому встановлена ОС Windows XP/Vista/7. Перед тим як підключати пристрій до комп'ютера необхідно перевірити апаратну частину (монтаж і підключення USB-кабелю), потім встановити програмне забезпечення, разом з яким буде встановлений і драйвер сканера, після чого сканер можна підключати до комп'ютера. У диспетчері пристроїв, в розділі *Інші пристрої*, ви також побачите підключений в системі сканер ISM-діапазону.

Кнопка *Export* призначена для експорту поточних відліків в файл *.csv*, який може бути завантажений в програму Exel для обробки і побудови графіків.

При запуску програми потрібно перевірити, що пристрій підключений: «Connected to Geoff's 2.4GHz Scanner». Повідомлення «Scanner not found» означає, що пристрій або не підключений, або не працює.

При запуску програми і підключенні сканера до USB в основному вікні програми ви побачите спектр сигналу діапазону 2,4 ГГц. Якщо поблизу немає бездротових пристроїв ISM-діапазону, спектр буде відображати фоновий шум і шум радіочастотної частини CYWUSB6935. Зверніть увагу, що вертикальна шкала (рівень сигналу) не відкалібрована, вона носить відносний характер, оскільки відображає рівень сигналу, який посилає радіомодулем. Спектр сигналу (рис. 15) відображає роботу безпроводової точки доступу Wi-Fi маршрутизатора (стандарту IEEE 802.11n), налаштованого на 8-й канал з центральною частотою 2,447 ГГц і розташованого на відстані у 10 м.

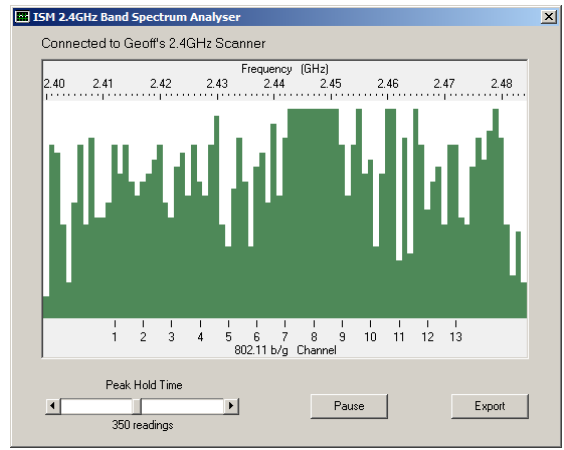

*Рис. 15. Спектр сигналу безпроводового маршрутизатора IEEE 802.11n*

## 8. ПОРІВНЯННЯ РОБОТИ АНАЛІЗАТОРІВ СПЕКТРУ

На рис. 16а і 16б представлено результати прийому при розташуванні аналізатора спектру в ближній зоні передавача для передача даних по безпроводовому каналу Bluetooth, а на рис. 16в і 16г — Wi-Fi діапазону 2,4 ГГц. З рисунків легко бачити, що суцільна збірка показала набагато якісніші результати через те, що в ній якісніше проведена збірка, використаний екран і зовнішня антена.

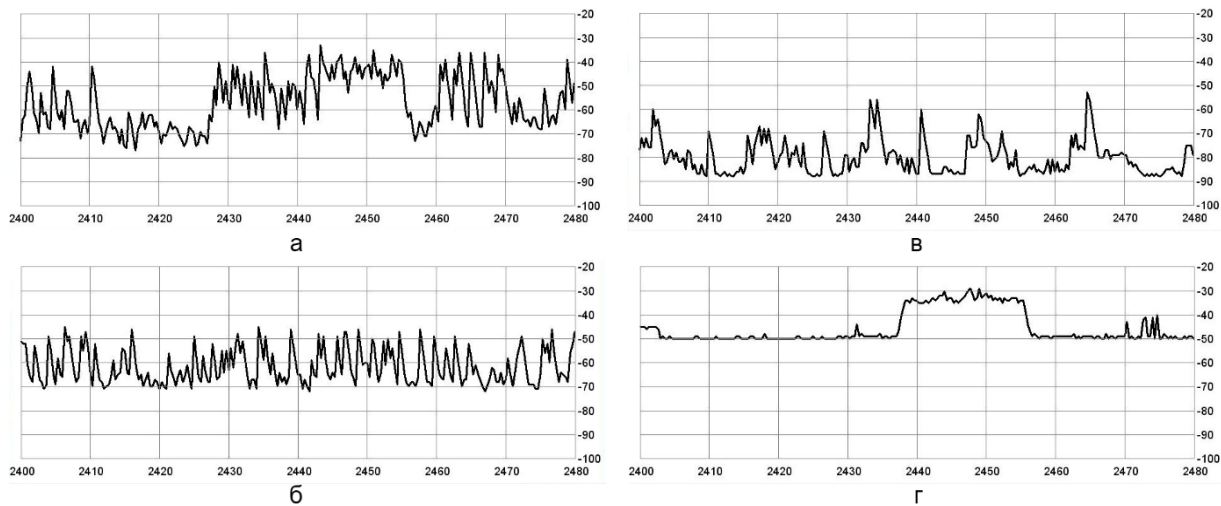

*Рис. 16. Приклади спектрів для різних збірок:*
*а — Bluetooth для модульної; б — Bluetooth для суцільної;*
*в — Wi-Fi для модульної; г — Wi-Fi для суцільної*

Після закінчення збірки, тестування і налагодження пристрої готові для використання як у технічний, так і у наукових задачах. Загальний вигляд усіх трьох пристроїв показаний на рис. 17.

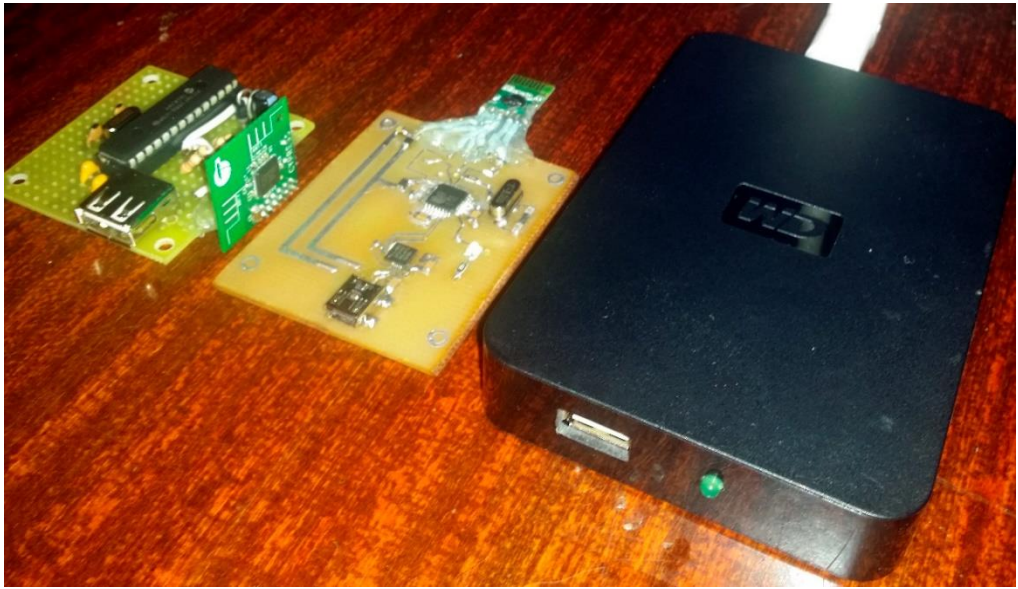

*Рис. 17. Загальний вигляд усіх приладів*

## 9. ВИСНОВКИ ТА ПЕРСПЕКТИВИ ПОДАЛЬШИХ ДОСЛІДЖЕНЬ

В статті представлені результати проектування і виготовлення аналізаторів спектру на готових компонентах (мікросхеми-трансивери для стандарту IEEE 802.15.4/ZigBee). Детально описаний процес проектування, виготовлення друкованих плат, збір приладів і програмування мікроконтролерів. Проведене тестування і внесення вдосконалень у існуючі прилади. При розводці плати виявлена залежність якості роботи приладу від якості його збірки, наявності електромагнітного екрану і типу антени.

В статті використане стороннє програмне забезпечення, а також розроблене на кафедрі інформаційної та кібернетичної безпеки програмне забезпечення для аналізу даних, зібраних с різних аналізаторів спектру.

Після детального тестування приладів і їх перевірки ми дійшли висновку, що можна виготовити більш компактне рішення для серійного випуску приладів.

Можливі напрямки подальших наукових досліджень включають більш глибокий статистичний аналіз, удосконалення підходів до вимірювання інформації і прогнозування. У майбутньому дані прилади можна інтегрувати в програмний комплекс ситуаційного центру, який консолідує в собі роботу з різними низькобюджетними моделями аналізаторів спектру заданого частотного діапазону.

**Volodymyr Yu. Sokolov**
MSc, senior lecturer
Borys Grinchenko Kyiv University, Kyiv, Ukraine
OrcID: 0000-0002-9349-7946
vladimir.y.sokolov@gmail.com

# COMPARISON OF POSSIBLE APPROACHES FOR THE DEVELOPMENT OF LOW-BUDGET SPECTRUM ANALYZERS FOR SENSORY NETWORKS IN THE RANGE OF 2.4–2.5 GHZ

**Abstract.** The article deals with the development, implementation and research of the spectrum analyzers that can be used in sensor networks and Internet systems of things. As an operating frequency range, 2.4–2.5 GHz ISM is selected. At the stage of hardware selection, a comparative analysis of existing available microcontrollers for the analysis of the spectrum, the choice of hardware interfaces, the ordering of the required modules and electrical components, as well as the input control is carried out. During development, several variants of spectrum analyzers on the basis of microcontroller and TI Chipcon CC2500 microcontrollers with USB interfaces, as well as Cypress CYWUSB6935 modules with LPT and USB interfaces, have been implemented. At the development stage, the development of the printed circuit board, its fabrication, component assembly, microcontroller programming, the verification of the assembly's robustness, making corrections, connecting to a personal computer and assembly in the case have been carried out. An analysis of existing software for collecting information on the state of the wireless broadcast is also conducted. According to the results of comparative experiments of various collections of spectrum analyzers, spectrographs for different types of signals were obtained. On these typical spectrographs a comparative analysis of the work of various prototypes was conducted. The offered approaches to building sensors on the basis of spectrum analyzers allow to create low-power modules for embedding in existing wireless information networks of enterprises for prevention of inter-channel interference and ensuring the integrity of data transmission. As a result of experiments, it is evident that not all existing modules are very different in characteristics, and the quality of their work is directly related to the type and quality of the antenna. The article gives the basic electric circuits, a list of elements, examples of PCBs, programmers, software and prototypes.

**Keywords:** spectrum analysis; wireless network; transceiver; ISM range.